\documentclass[prd,twocolumn,showpacs,showkeys]{revtex4}
\usepackage{graphicx}% Include figure files
\usepackage{dcolumn}% Align table columns on decimal point
\usepackage{bm}% bold math

\begin{document}

\title{Structure formation  in the quasispherical Szekeres model}

\author{Krzysztof  Bolejko}
\affiliation{Nicolaus Copernicus Astronomical Center, Polish Academy of Sciences,
 ul. Bartycka 18, 00-716 Warsaw, Poland}
\email{bolejko@camk.edu.pl}
\homepage{http://www.camk.edu.pl/~bolejko}

\date{\today}

\begin{abstract}
Structure formation in the Szekeres model is investigated. Since the Szekeres
model is an inhomogeneous model with no symmetries, it is possible to examine
the interaction of neighboring structures and its impact on the growth of a
density contrast. It has been found that the mass flow from voids to  clusters
enhances the growth of the density contrast. In the model presented here, the
growth of the density contrast is almost 8 times faster than in the linear
approach.
\end{abstract}


\maketitle

\section{Introduction}

Galaxies redshift surveys  indicate that our Universe is inhomogeneous.
Galaxies form structures like clusters, superclusters, and voids. The most
popular methods which are used to describe the evolution of these structures
are N-body simulations (\cite{NB1,NB2,NB3}) and the linear approach. However, because the
present day density contrast is large the linear approach is in most cases
inadequate. 
On the other hand the N-body simulations  
 describe  the evolution of large amount of particles which interact gravitationally.  
However, interactions between particles are described by the Newtonian mechanics.  
In Newtonian mechanics matter does not affect light propagation, 
hence within the N-body simulations it is impossible to estimate the influence 
of matter distribution on light propagation. 
In general relativity the situation is diffrent, the geometry defined by
matter distribution tells the light along which paths to propagate.
Thus, in order to have a suitable model which would predict
a proper evolution of the  density contrast and be adequate to trace  light
propagation  models based on  exact solutions of the Einstein equations need
to be used. In this paper the  Szekeres model is employed to study the
evolution of a galaxy supercluster and an adjourning void. The Szekeres model
is an exact solution of the Einstein equations, which is inhomogeneous and has
no symmetries. Being an exact model of spacetime geometry, the Szekeres model
can be adopted not only to describe the evolution of cosmic structures but
also to examine light propagation.

The structure of this paper is as follows: Sec. \ref{szekmdl} presents the
Szekeres model; in Sec. \ref{vpc} the model of the double-structure is
presented; in Sec \ref{evol} the evolution a void and an adjourning cluster is
presented. The results of this evolution are compared with the results
obtained in the linear approach and in the inhomogeneous spherically symmetric
Lema\^itre--Tolman model.

\section{The Szekeres model}\label{szekmdl}

The metric of the Szekeres  \cite{Sz1} models is of the following form:

\begin{equation}
{\rm d} s^2 = {\rm d} t^2 - {\rm e}^{2 \alpha} {\rm d} z^2 - {\rm e}^{2 \beta}
( {\rm d}x^2 + {\rm d}y^2). \label{ds2}
\end{equation}

The components of the metric are as follows:

\begin{eqnarray}
{\rm e}^{\beta} &=& \Phi(t,z) {\rm e}^{\nu (x,y,z)}, \\
{\rm e}^{\alpha} &=& h(z) \Phi(t,z) \beta,_z,
\end{eqnarray}

where $h(z)$ is an arbitrary function of $z$, and ${\rm e}^{-\nu}$ is:

\begin{equation}
 {\rm e}^{-\nu} = A(z)(x^2 + y^2) + 2B_1(z)x + 2 B_2(z)y + C(z).
 \label{e-n}
 \end{equation}

The functions $A(z), B_1(z), B_2(z), C(z)$ are not independent but obey the
following relation:

\begin{equation}
C(z)   =  \frac{B_1{}^2(z)}{A(z)} + \frac{B_2{}^2(z)}{A(z)} + \frac{1}{4A(z)}
\left[ \frac{1}{h^2(z)} + k(z) \right]
\end{equation}

The Einstein equations reduce to following two:

\begin{equation}
\Phi,_t^2 (t,z) = \frac{2M(z)}{\Phi(t,z)} - k(z) + \frac{1}{3} \Lambda
\Phi^2(t,z), \label{vel}
\end{equation}

\begin{equation}
\kappa \epsilon = \frac{ \left( 2 M {\rm e}^{3 \nu} \right),_z}{{\rm e}^{2
\beta} {\rm e}^{\beta},_z}. \label{rho}
\end{equation}

In a Newtonian limit $M c^2/G$ is equal to the mass inside the shell of 
radial coordinate $z$.  However, it is not an integrated rest mass but 
active gravitational mass that generates a gravitational field.

Eq. (\ref{vel}) can be integrated:

\begin{equation}
\int_0^{\Phi}\frac{{\rm d}
\tilde{\Phi}}{\sqrt{\frac{2M(z)}{\tilde{\Phi}} - k(z) + \frac{1}{3} \Lambda
\Phi^2}} = c \left[t- t_B(z)\right], \label{cal}
\end{equation}
where $t_B$ appears as an integration constant, and is an arbitrary function of
$z$. This means that the Big Bang is not a single event as in the Friedmann
models, but occurs at different times for different distances from the origin.

As can be seen the Szekeres model is specified by 6 functions. However, by a
choice of the coordinates, the number of independent functions can be reduced
to 5.

The Szekeres model is known to have no symmetry (Bonnor, Sulaiman and Tomimura 
\cite{BST}). It is of great flexibility and wide application in cosmology 
(Bonnor and Tomimura  \cite{BT}), and in astrophysics
(Szekeres \cite{Sz2}; Hellaby and Krasi\'nski \cite{HK}), and still it can
be used as a model of many astronomical phenomena. This paper aims to present
the application of the Szekeres model to the process of structure formation.

\subsection{Coordinate system}

The coordinate system in which the metric is of form (\ref{ds2}) can be
interpreted as a stereographic projection of polar coordinates. This can be
seen if the following transformation is considered:

\begin{eqnarray}
&& A = \frac{1}{2S}, \nonumber \\
&& B_1 = - \frac{P}{2S}, \nonumber \\
&& B_2 = - \frac{Q}{2S}, \nonumber \\
&& C = \frac{P^2}{2S} +  \frac{Q^2}{2S} + \frac{S}{2} = \frac{B_1^2}{A} +
\frac{B_2^2}{A} + \frac{\varepsilon}{4A}, \nonumber \\
&& \varepsilon =  \frac{1}{h^2} + k.
\label{sp1}
\end{eqnarray}

After this transformation we obtain:

\begin{eqnarray}
&& {\rm e}^{ 2 \nu} (dx^2 + dy^2) = \nonumber \\
&&=  \frac{(dx^2 + dy^2)}{   \left[ A (x^2 + y^2) +
2B_1 x + 2B_2 y + C \right]^2  }  = \nonumber \\
&& =  \frac{(dx^2 + dy^2)}{   \left[ \frac{1}{2S}(x^2+y^2) - 2 \frac{P }{2S} x 
- 2 \frac{Q}{2S} y +  \frac{ P^2}{2S} +  \frac{ Q^2}{2S} + 
\frac{S}{2} \right]^2}  = \nonumber \\
&& =  \frac{(dx^2 + dy^2)}{  \frac{S^2 }{4} \left[ \left(\frac{x-P}{S} 
\right)^2 + \left(\frac{y-Q}{S} \right)^2 + \varepsilon \right]^2  }.
\end{eqnarray}

When $\varepsilon = 1$ the above transformation is the stereographic
projection of a sphere, when $\varepsilon = 0$ the surface is a plane, and when
$\varepsilon = -1$ it is the stereographic projection of a hyperboloid.

As we are interested in the Friedmann limit of our model, i.e. we expect it
becomes an homogeneous Friedmann model in a large distance from the origin, we
will focus only on the  $\varepsilon = 1$ case.

Then the transformation of the following form:

 \begin{eqnarray}
 x - P &=& S {\rm cot} \left( \frac{ \theta }{2} \right) \cos (\phi) \nonumber
 \\
 y - Q &=& S {\rm cot} \left( \frac{ \theta }{2} \right) \sin (\phi) \nonumber
 \\
 z &=&  r
 \label{sphtrsf}
 \end{eqnarray}

leads to:

\begin{equation}
{\rm e}^{2 \beta} (dx^2 + dy^2) = \Phi^2 \left( {\rm d}{\theta}^2 + \sin^2
\theta {\rm d}{\phi}^2 \right).
\end{equation}

After transformation (\ref{sp1}) and (\ref{sphtrsf}) the metric (\ref{ds2})
becomes:

\begin{eqnarray}
&& {\rm d} s^2 = c {\rm d} t^2 - \left\{ \frac{( \Phi,_r + \Phi \nu,_r)^2}{1 -k}
+ \Phi^2 {\rm e}^{2 \nu} \left[ S,_r^2 \cot^2 \frac{\theta}{2} \right. \right.
 \nonumber \\ && \left. \left. + 2 S,_r \cot
\frac{\theta}{2} \left( Q,_r \sin \phi + P,_r \cos \phi \right)  +  \left(
P,_r^2 + Q,_r^2 \right) \right] \right\} {\rm d} r^2  \nonumber \\
&& -  \Phi^2 {\rm e}^{2 \nu} \left[ 2 S \cot \frac{\theta}{2}  \left( Q,_r
\cos \phi - P,_r \sin \phi \right)  \right] {\rm d} r {\rm d} 
{\phi} \nonumber \\ 
&&  +  2\Phi^2 {\rm e}^{\nu} \left(   Q,_r \sin \phi + P,_r \cos \phi 
+ S,_r \cot \frac{\theta}{2} \right) {\rm d} r {\rm d} {\theta} \nonumber \\
&& -  \Phi^2  \left( {\rm d}{\theta}^2 + \sin^2\theta {\rm d} {\phi}^2 \right),
\label{ds2ss}
\end{eqnarray}
where:

\begin{equation}
{\rm e}^{\nu} = \frac{1 - \cos \theta }{S},
\label{enu}
\end{equation}

and:

\begin{equation}
 \nu,_r = \frac{S,_r \cos \theta + \sin \theta \left( P,_r \cos \phi + Q,_r
\sin \phi \right)}{S}. \label{nuz}
\end{equation}

As can be seen, if $t$ = const, and $r$ = const, the above metric becomes the
metric of the 2 dimensional sphere. Hence,  every $t$ = const and $r$ = const
slices of the Szekeres $\varepsilon = 1$ space-time is a sphere. Therefore,
the $\varepsilon =1$ case is often called quasispherical model. However, as
$S, P$ and $Q$ are now functions of $r$, spheres are not concentric. For the
spheres to be concentric, the following conditions must hold:

\begin{eqnarray}
P,_r = 0, \nonumber \\
Q,_r = 0, \nonumber \\
S,_r = 0.
\label{sscon}
\end{eqnarray}
Such conditions lead to spherical symmetric case, and the metric (\ref{ds2ss})
becomes the line element of the Lema\^itre--Tolman model \cite{Lem,Tol}.

\subsection{The Friedmann limit}\label{frdlmt}

The Friedmann limit is an essential element of our model. The model presented
in this paper describes the evolution of a void with an adjourning cluster in
the expanding Universe. Far away from the origin  density and velocity
distributions tend to the values that they would have in a Friedmann model.
Consequently the values of the time instants and values of the density  and
velocity fluctuations are calculated with respect to this homogeneous
background.

 The Friedmann limit follows when:

\begin{eqnarray}
\Phi(r,t) &=& R(r) f(t), \\
 k(r) &=& -k_0 R^2(r),
\end{eqnarray}
where $k_0$ is the curvature index of  the FLRW models.

The above conditions are sufficient to obtain the homogeneous FLRW model, and
the metric  (\ref{ds2}) assumes the Goode and Wainwright \cite{GW} form of the
FLRW model. Then from Eq. (\ref{rho}) follows:

\begin{equation}
 M(r) = M_0 R^3(r),
\end{equation}
where $M_0$, expressed by FLRW parameters is $M_0 = (1/2) (\Omega_m
H_0^2/c^2)$. Inserting the above into Eq. (\ref{cal}) it follows that $
t_B(r)$ = const, which implies that the Big Bang was simultaneous. Although,
the metric in polar coordinates (\ref{ds2ss}) is still  not diagonal, under
 the transformation (\ref{sscon}), the metric obtains a more usual form:

\begin{equation}
{\rm d} s^2 = {\rm d} t^2 - \frac{f^2(t)}{1 - k_0 R^2} {\rm d} R^2 - R^2
f^2(t) {\rm d} \Omega^2 ,
\end{equation}
where
${\rm d} \Omega^2 =  {\rm d}{\theta}^2 + \sin^2\theta {\rm d} {\phi}^2.$

\section{Structure formation}\label{strfr}

This paper aims to present the application of the Szekeres model to the
process of structure formation. In this section  the model of an evolving void
with adjourning supercluster is presented. As will be seen the use of Szekeres
model gives better understanding of the structure formation, and shows the
importance of voids in process of cluster formation.

The model is expected to remain consistent with the  astronomical data. As
mentioned in Sec. \ref{frdlmt}  the density fluctuations, as well as time
instants are calculated with respect to the homogeneous background model. The
chosen background model is the FLRW model with the density:

 \begin{equation}
 \rho_b = \Omega_m \times \rho_{cr} = 0.3 \times \frac{3H_0^2}{8 \pi G}.
  \label{rbdf}
  \end{equation}
The Hubble constant is of   $H_0 =72$ km $^{-1}$ Mpc$^{-1}$, and the
cosmological constant corresponds to $\Omega_{\Lambda} = 0.7$, 
 where $\Omega_{\Lambda} = (1/3)  ( c^2 \Lambda/H_0^2)$.

 Below the density distribution and the evolution of a void and an adjourning
 cluster is calculated. It can be seen  from Eqs. (\ref{rho}), (\ref{enu}), 
 and (\ref{nuz}) that to calculate the density distribution for any instant
$t_i$, one needs to now 5 functions: $M(r), S(r), Q(r), P(r)$, and 
$\Phi(t_i,r)$. The explicit forms of these functions are presented 
below. Using these functions the density distribution of the present day 
structures can be calculated (see Sec. \ref{vpc}). Then, the evolution of 
the system can be traced back in time. 
The density distribution depends on time only via the function $\Phi(t,r)$ and 
its derivative. The value of the $\Phi(t,r)$ for any instant can be calculated
by solving the differential equation [see Eq. (\ref{vel})]. 
In most cases, as in this paper, this equation can be solved only numerically. 
To solve this equation one needs to know the initial conditions: $\Phi(t_0,r)$, 
and the functions $M(r)$, and $k(r)$ as well as the value of $\Lambda$. 
This equation was solved numerically using the fourth-order Runge--Kutta 
method \cite{NR}. Knowing the value of $\Phi(t,r)$ for any instant the density 
distribution can be calculated as described above.

\subsection{Observational constrains}

Astronomical observations show that in small scales matter distribution and
expansion of the space are not homogeneous. The measurements of  matter
distribution imply that density varies from $\rho \approx 0.06 \rho_b$ in
voids \cite{Hoy} to  $\rho$ equal several tens of background
density ($\rho_b$) in clusters  \cite{Bar}. These structures are of
diameters from several Mpc up to several tens of Mpc. However, if the
averaging is considered on large scales, the density varies from $0.3 \rho_b$
to $4.4 \rho_b$ \cite{Kol, Hud},  and the structures
are of several tens of Mpc.

\subsection{Model of a void with an adjourning supercluster}\label{vpc}

As mentioned above, to specify the model one needs to know 5
functions of the radial coordinate. Let us define the radial coordinate as a
current value of $\Phi$:
\begin{equation}
r:= \Phi(z,t_0)
\end{equation}

\begin{figure}
    \begin{center}
    \includegraphics[scale=0.24]{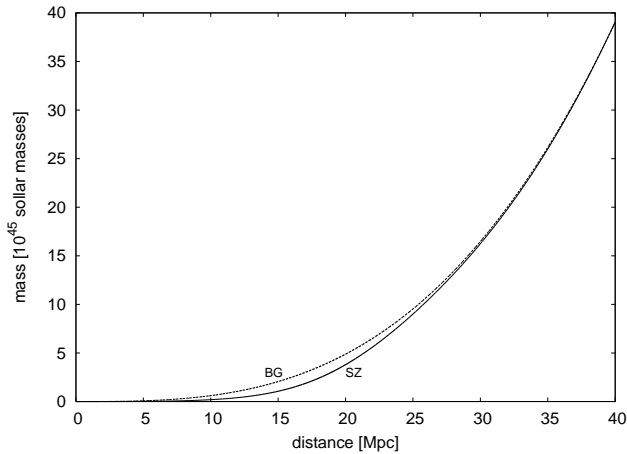}
\caption{The mass distribution within the homogeneous background (BG) and in
the Szekeres model (SZ).}
    \label{fig1}
    \end{center}
\end{figure}

Three out of these five unknown functions can be $P(r), Q(r), S(r)$. However
the physically important quantities are not these functions, but their
gradients. If $P(r), Q(r), S(r)$ are constant, then as can be seen from Eqs.
(\ref{nuz}), (\ref{enu}) and Eq. (\ref{rho}), the density distribution and the
evolution Eq. (\ref{vel}) do not depend on them. Then the Szekeres model
becomes the  Lem\^itre--Tolman model. The explicit forms of these functions
are presented in next subsections. The next two functions can be either,
$t_B(r), M(r), k(r)$ or any other combination of functions, from which these
can be calculated. The function $M(r)$ describes the active gravitational mass
inside the $t = $ const, $r =$ const sphere. The assumed mass distribution is
presented in Fig. \ref{fig1}. The void is placed at the
origin, so the mass of the model in Fig. \ref{fig1} is below the background mass,
 but then it is compensated by more dense regions, and soon, at the distance of
about $30$ Mpc, the mass distribution becomes goes over into the homogeneous
background. To define the model we need one more function. Let us assume that
the bang time function is constant and equal to zero. Then from Eq.
(\ref{cal}) the function $k(r)$ can be calculated.

\subsubsection{Model 1}\label{model1}

As mentioned above, if the functions $P(r), Q(r)$ and $S(r)$ are constant, the
quasispherical Szekeres model becomes a Lema\^itre--Tolman model. Let us then
consider the simplest generalistion of the Lema\^itre--Tolman model. Let us
focus on a model  with $S(r)$ and $P(r)$ being constant, and $Q(r)$ being
chosen as below.

Let us choose:

\begin{eqnarray}
S &=& 140, \nonumber \\
P &=& 10, \nonumber \\
Q &=& -113 \ln (1+r).
\end{eqnarray}

\begin{figure*}
    \includegraphics[scale=0.26]{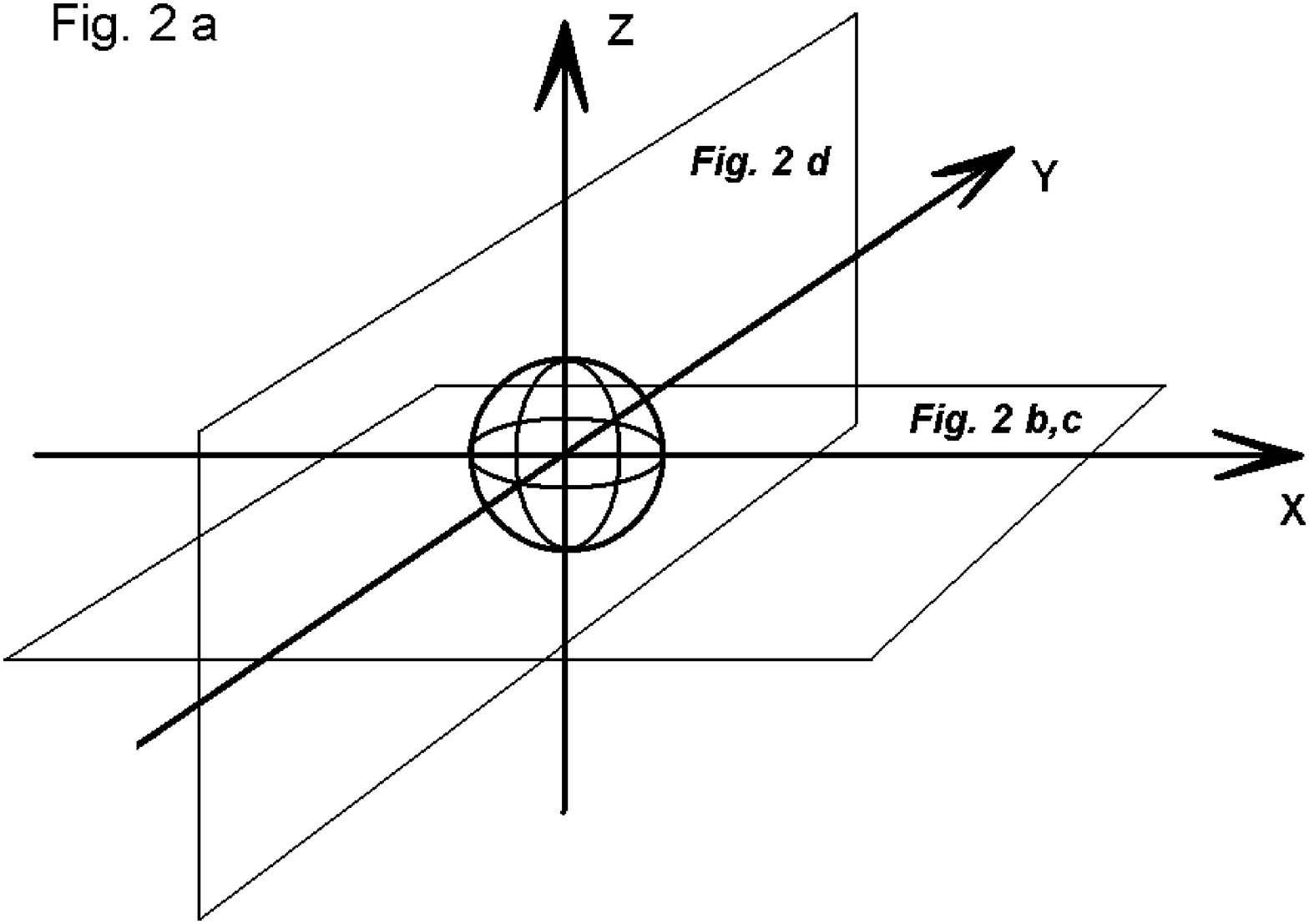}
    \includegraphics[scale=0.25]{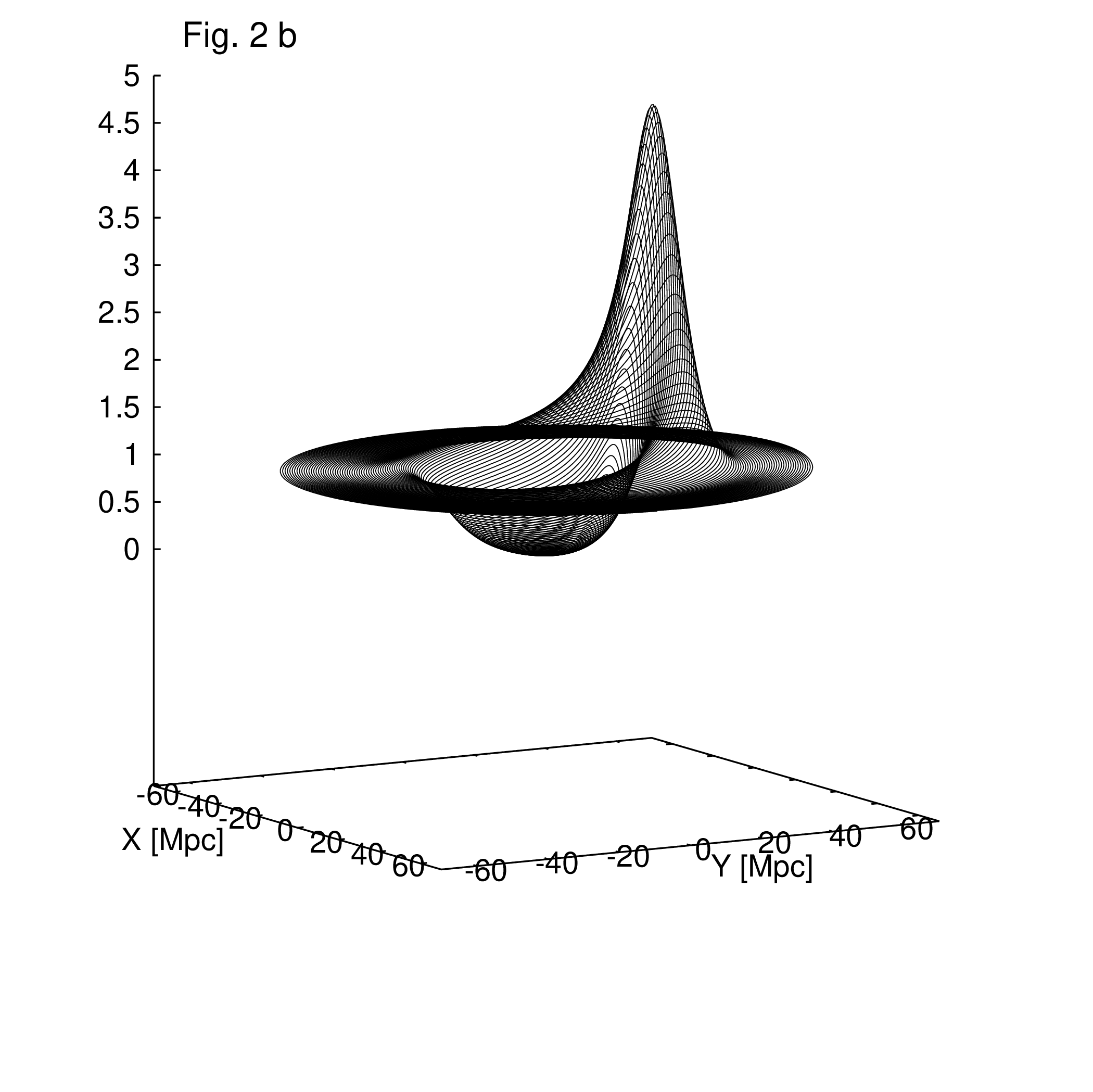}
%%%%%%%%%%%%%%%%%%%%%%%%%%%%%%%%%%%%%%%%%%
% Colour figs - online only:
%   \includegraphics[scale=0.23]{fig2c_col.eps}
%   \includegraphics[scale=0.23]{fig2d_col.eps}
%%%%%%%%%%%%%%%%%%%%%%%%%%%%%%%%%%%%%%%%%%   
% grayscale images - to be published 
    \includegraphics[scale=0.23]{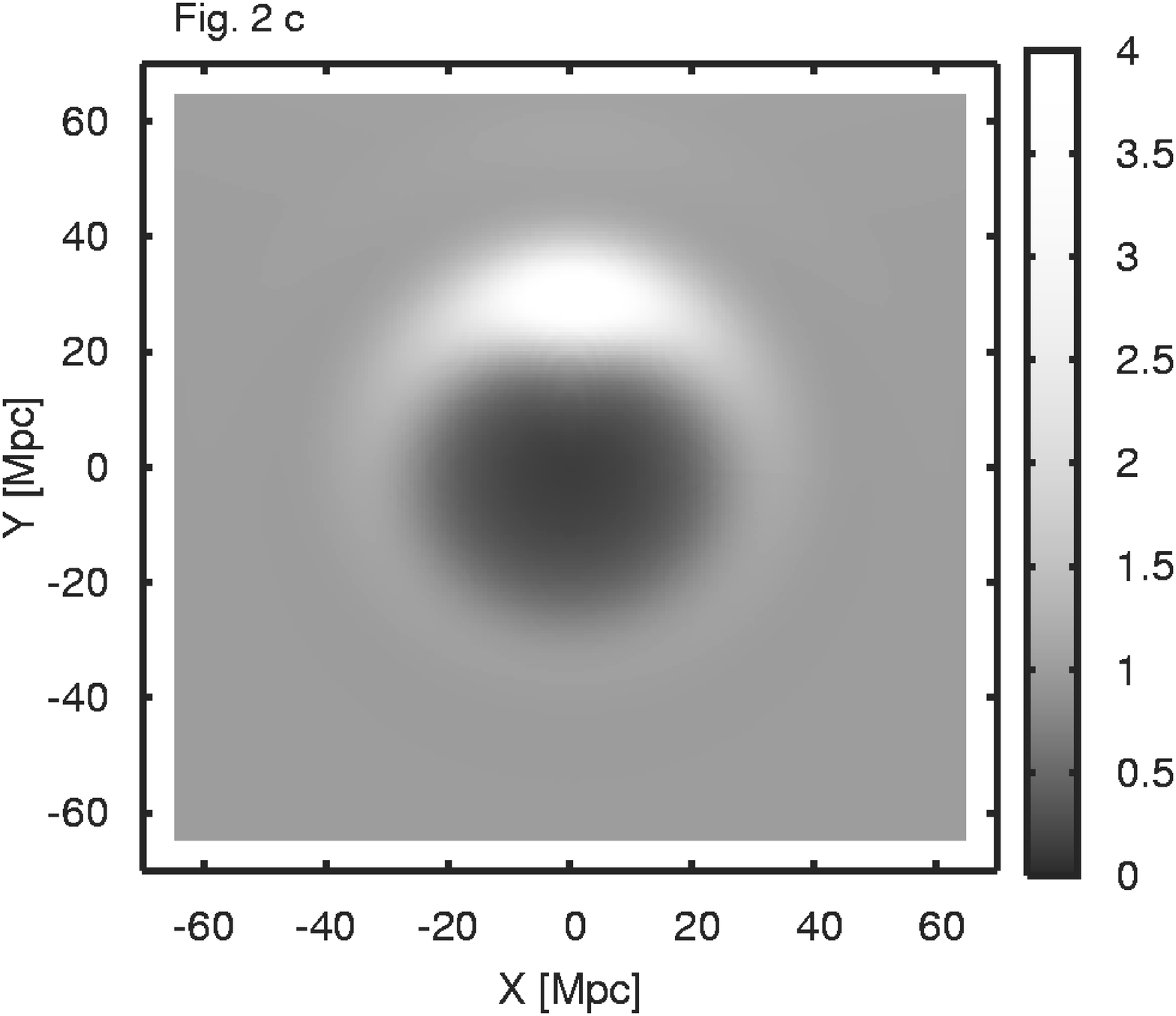}
    \includegraphics[scale=0.23]{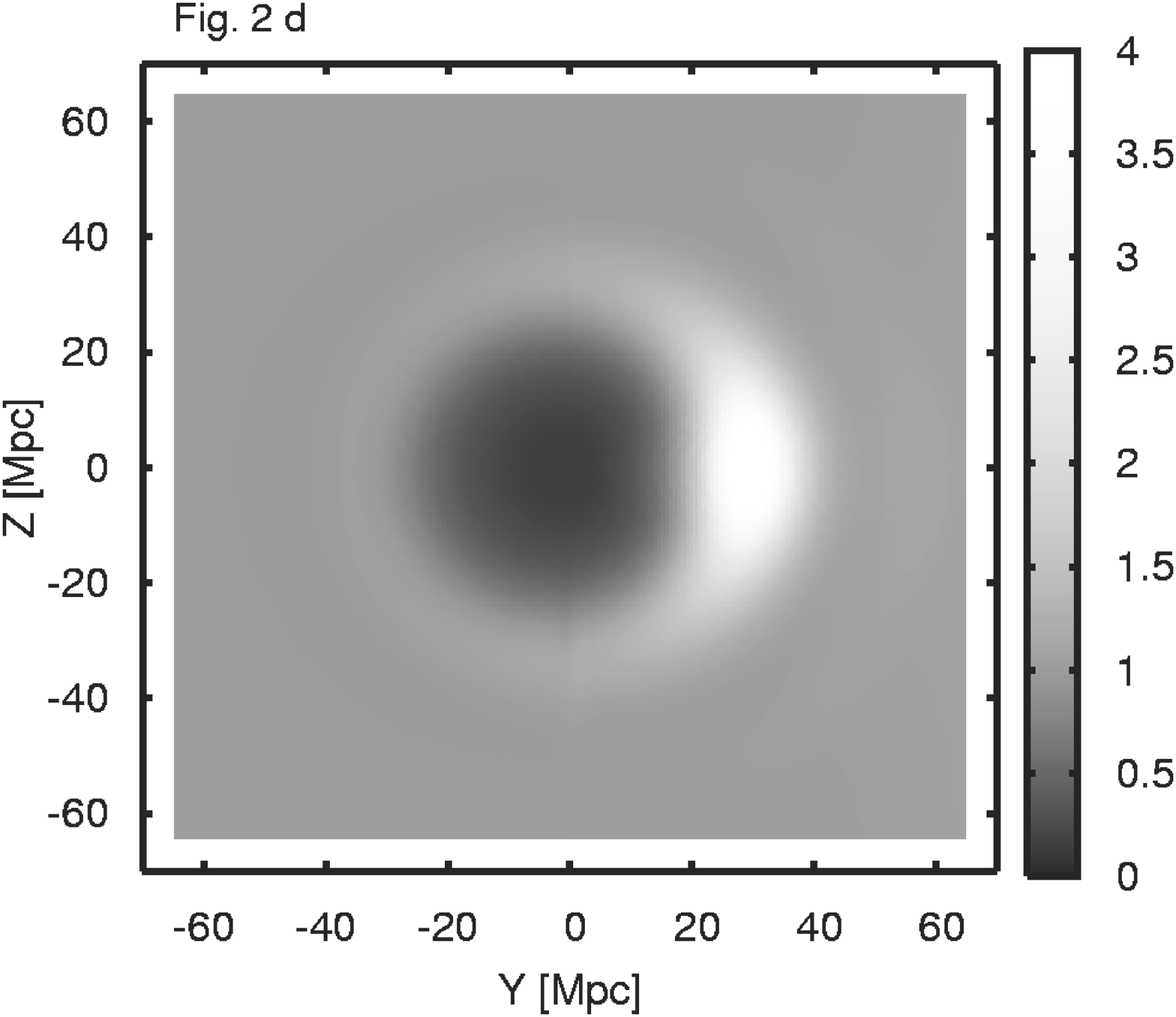}
%%%%%%%%%%%%%%%%%%%%%%%%%%%%%%%%%%%%%%%%%5
    \caption{
    ("to editor: for online version please use color figures:
	fig2c\_col.eps and fig2d\_col.eps")
    The present day density distribution of model 1 (Sec. 
    \ref{model1}). Fig. 2(a) presents a schematic view. 
    Figs. 2(b) -- 2(d) presents the density
     distribution in background units. Coordinates $X,Y,Z$ are defined as 
     follows: $X = \Phi(t_0,r) \sin \theta \cos \phi$,   
    $Y = \Phi(t_0,r) \sin \theta \sin \phi$, $Z = \Phi(t_0,r) \cos \theta$.} 
    \label{fig2}
\end{figure*}

The density distribution was calculated from Eq. (\ref{rho}), and it is
presented in Figs. \ref{fig2}(a) ---  \ref{fig2}(d). Fig.  \ref{fig2}(a) 
presents a schematic view of the
structure. Fig.  \ref{fig2}(b) and Fig.  \ref{fig2}(c) depict
 the horizontal  cross section through the
equator ($Z = 0$), so it goes through the void and the cluster [as presented
in Fig. \ref{fig2}(a)]. Fig  \ref{fig2}(d) depicts the vertical ($X = 0$) cross 
section, so it goes through the void and the cluster 
[as presented in Fig. \ref{fig2}(a)].

It should be stressed that the shapes presented in Figs. 
\ref{fig2}(b)  ---  \ref{fig2}(d), are a bit distorted in comparison with 
the real density distribution. The Szekeres model describes the density
distribution in a curved space, and it is impossible to map it into 
a 2 dimensional flat surface (such as a sheet of this paper).
 This problem is similar to drawing maps of our globe.

\subsubsection{Model 2}\label{model2}

Let us consider the following functions:

\begin{eqnarray}
S &=& - r^{0.59}, \nonumber \\
P &=& 0.83 \times r^{0.59}, \nonumber \\
Q &=& 0.4 \times r^{0.59}.
\end{eqnarray}

\begin{figure*}
    \includegraphics[scale=0.21]{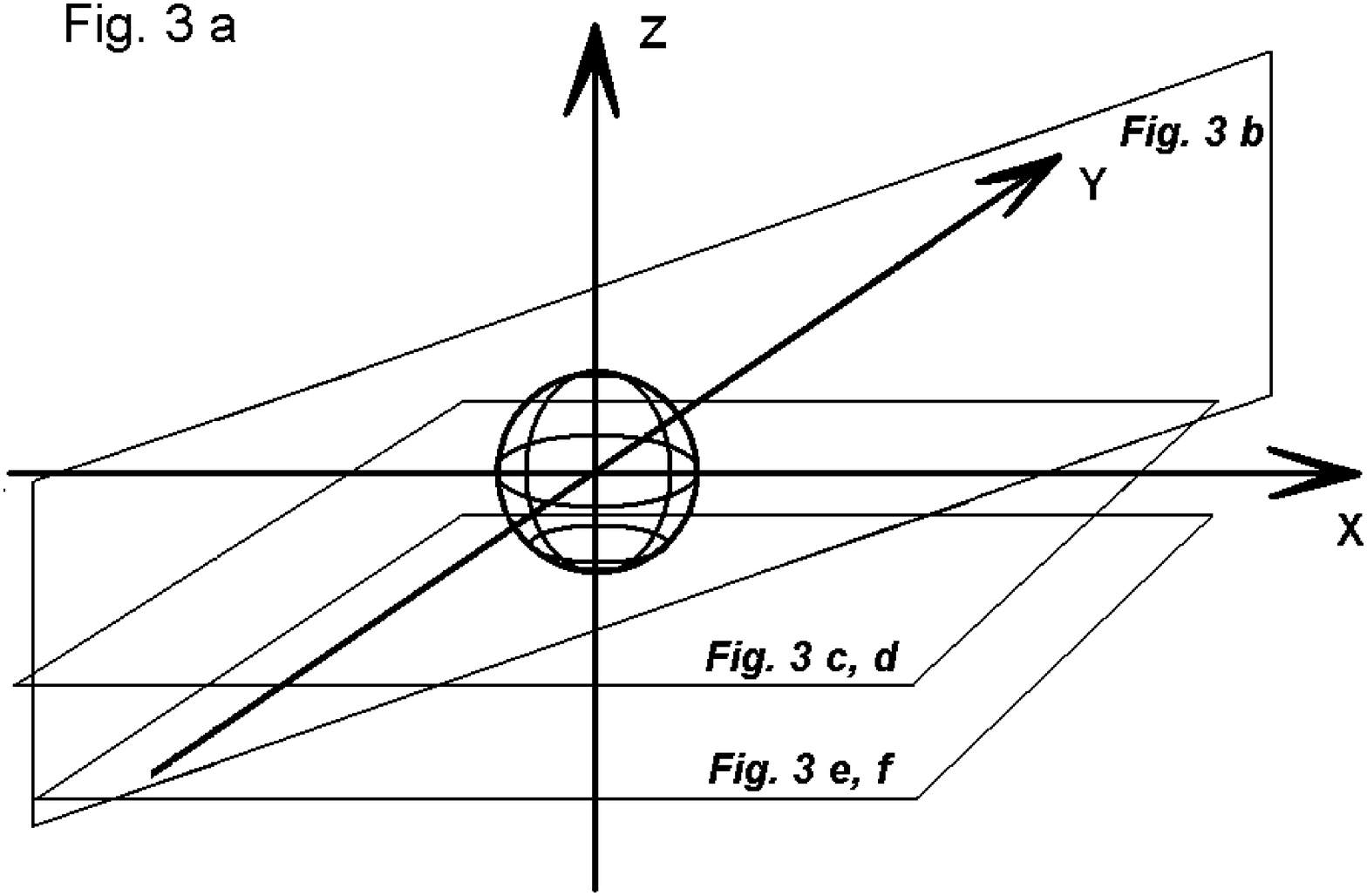}
%%%%%%%%%%%%%%%%%%%%%%%%%%%%%%%%%%%%%%%%%%
% Colour figs - online only:
%   \includegraphics[scale=0.23]{fig3b_col.eps}
%%%%%%%%%%%%%%%%%%%%%%%%%%%%%%%%%%%%%%%%%%   
% grayscale images - to be published:
  \includegraphics[scale=0.2]{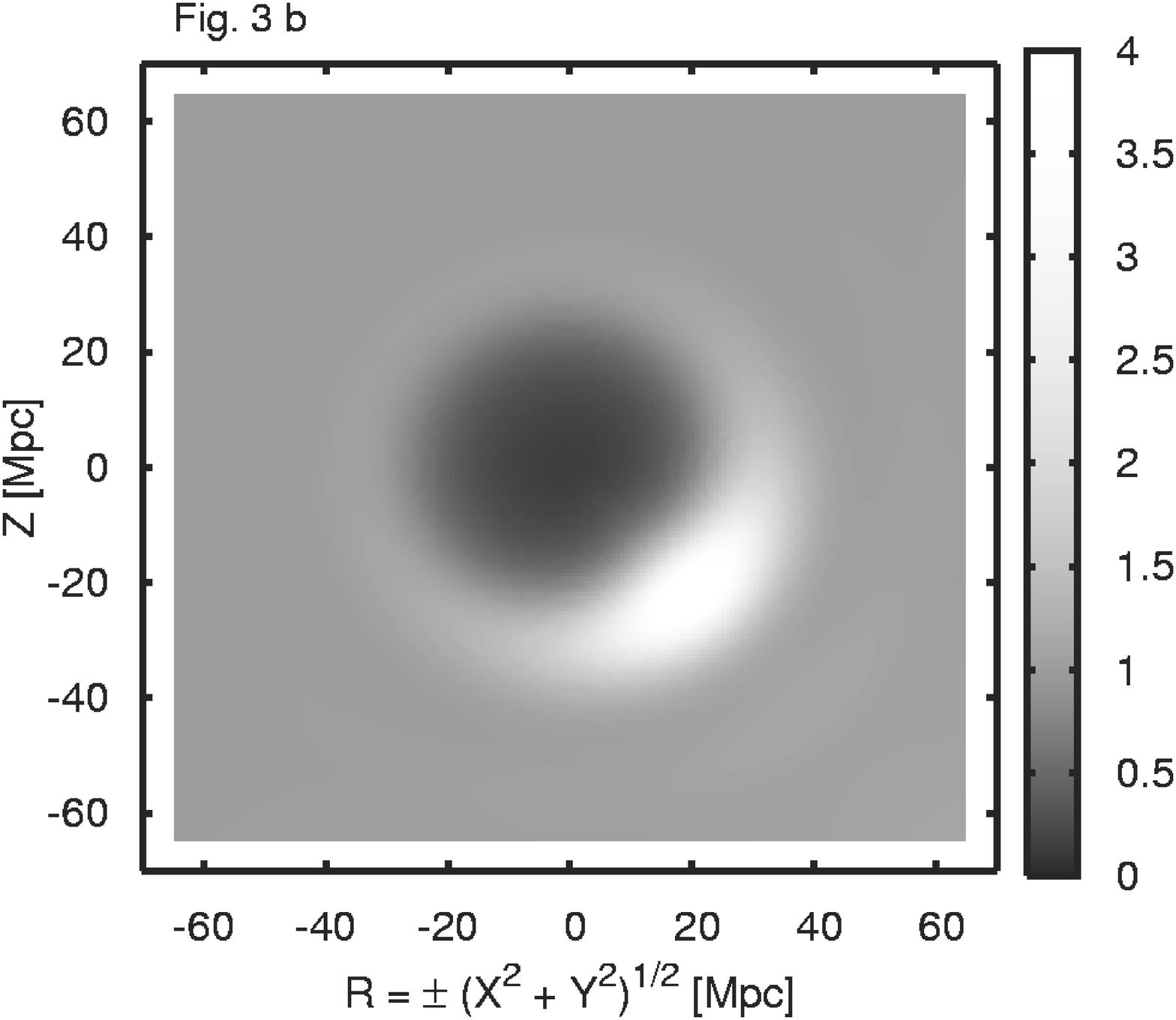}
%%%%%%%%%%%%%%%%%%%%%%%%%%%%%%%%%%%%%%%%%%       
  \includegraphics[scale=0.2]{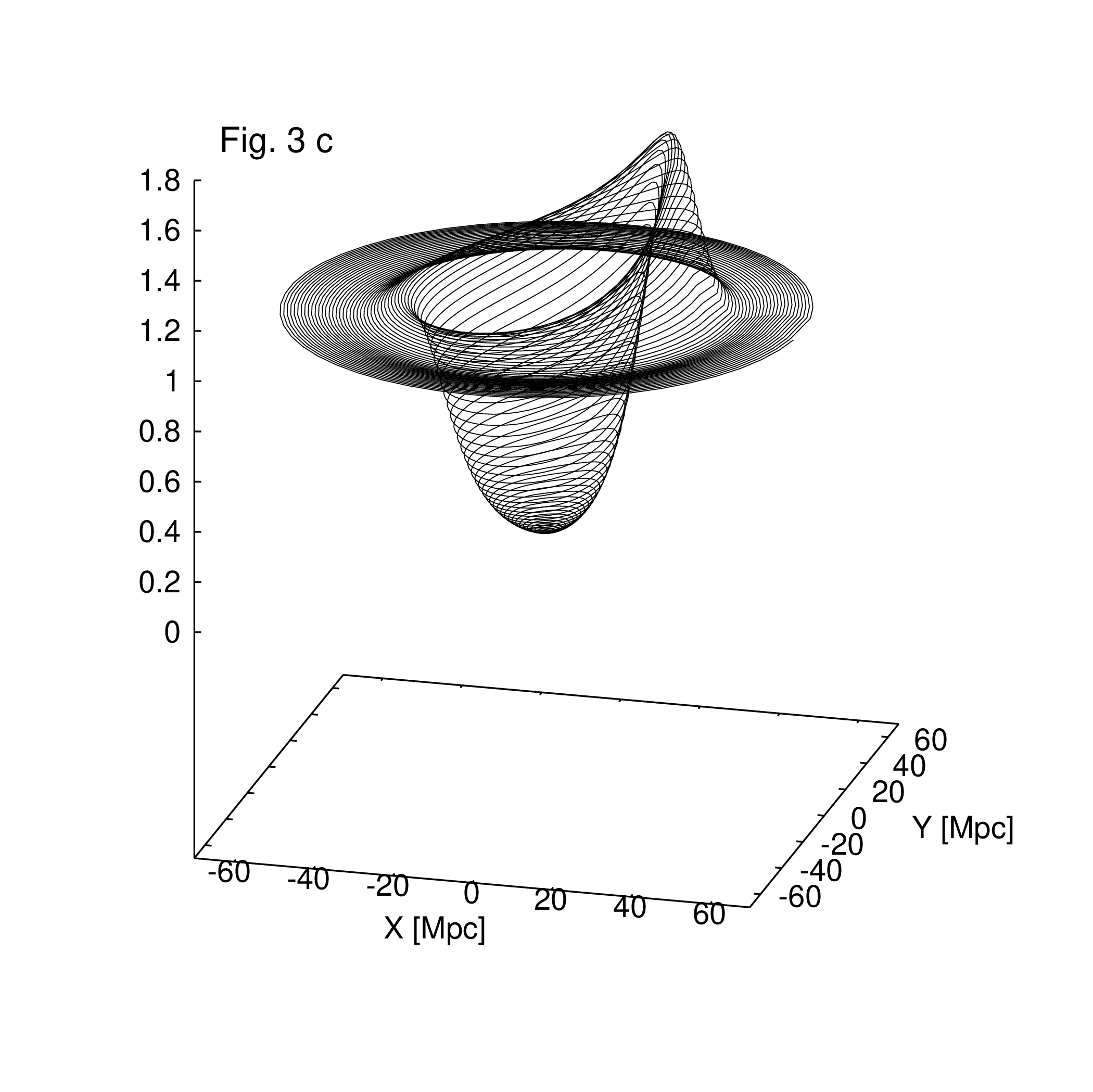}
%%%%%%%%%%%%%%%%%%%%%%%%%%%%%%%%%%%%%%%%%%       
% Colour figs - online only:
%   \includegraphics[scale=0.23]{fig3d_col.eps}
%%%%%%%%%%%%%%%%%%%%%%%%%%%%%%%%%%%%%%%%%%       
% grayscale images - to be published:
   \includegraphics[scale=0.2]{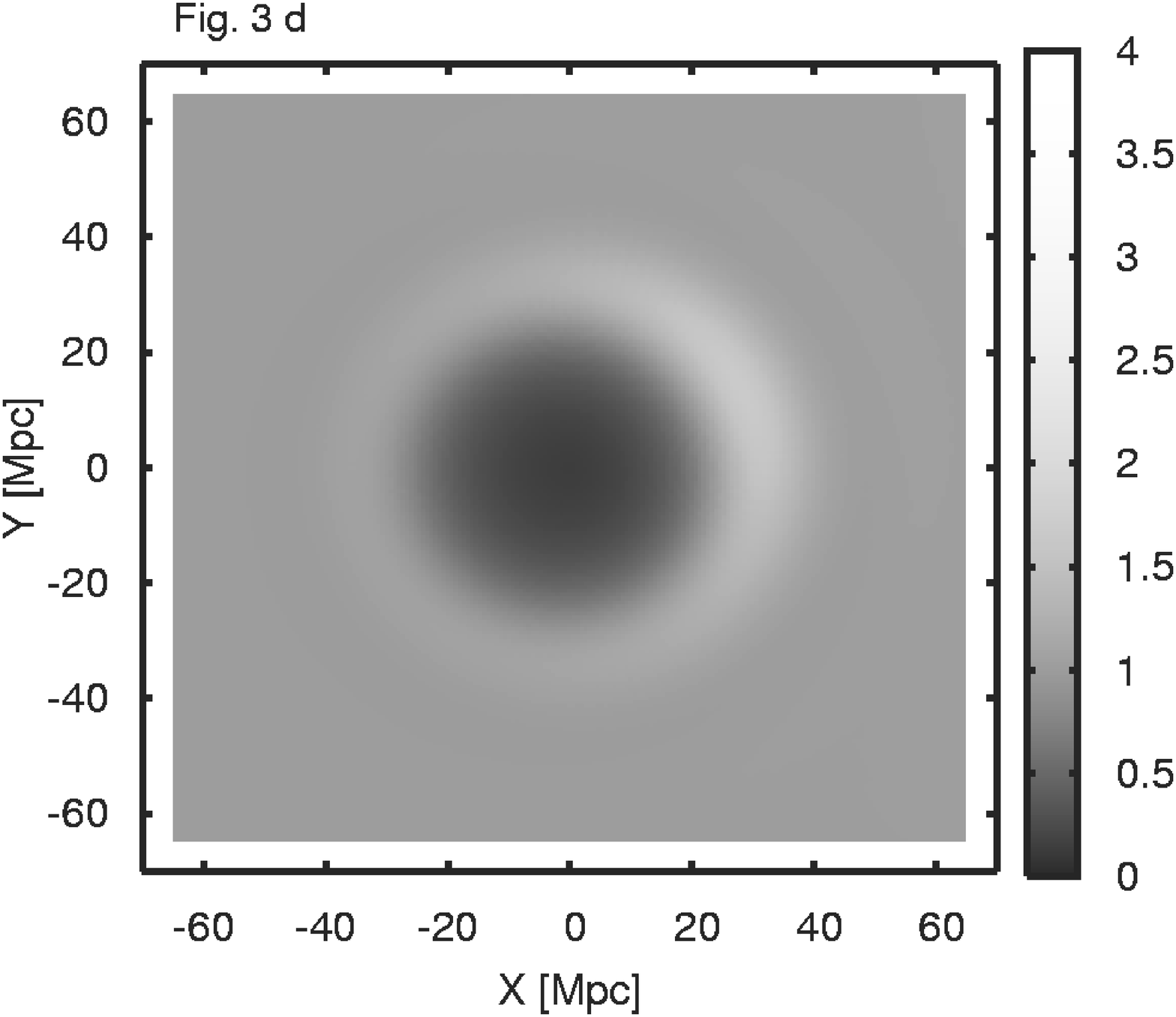}
%%%%%%%%%%%%%%%%%%%%%%%%%%%%%%%%%%%%%%%%%%       
   \includegraphics[scale=0.2]{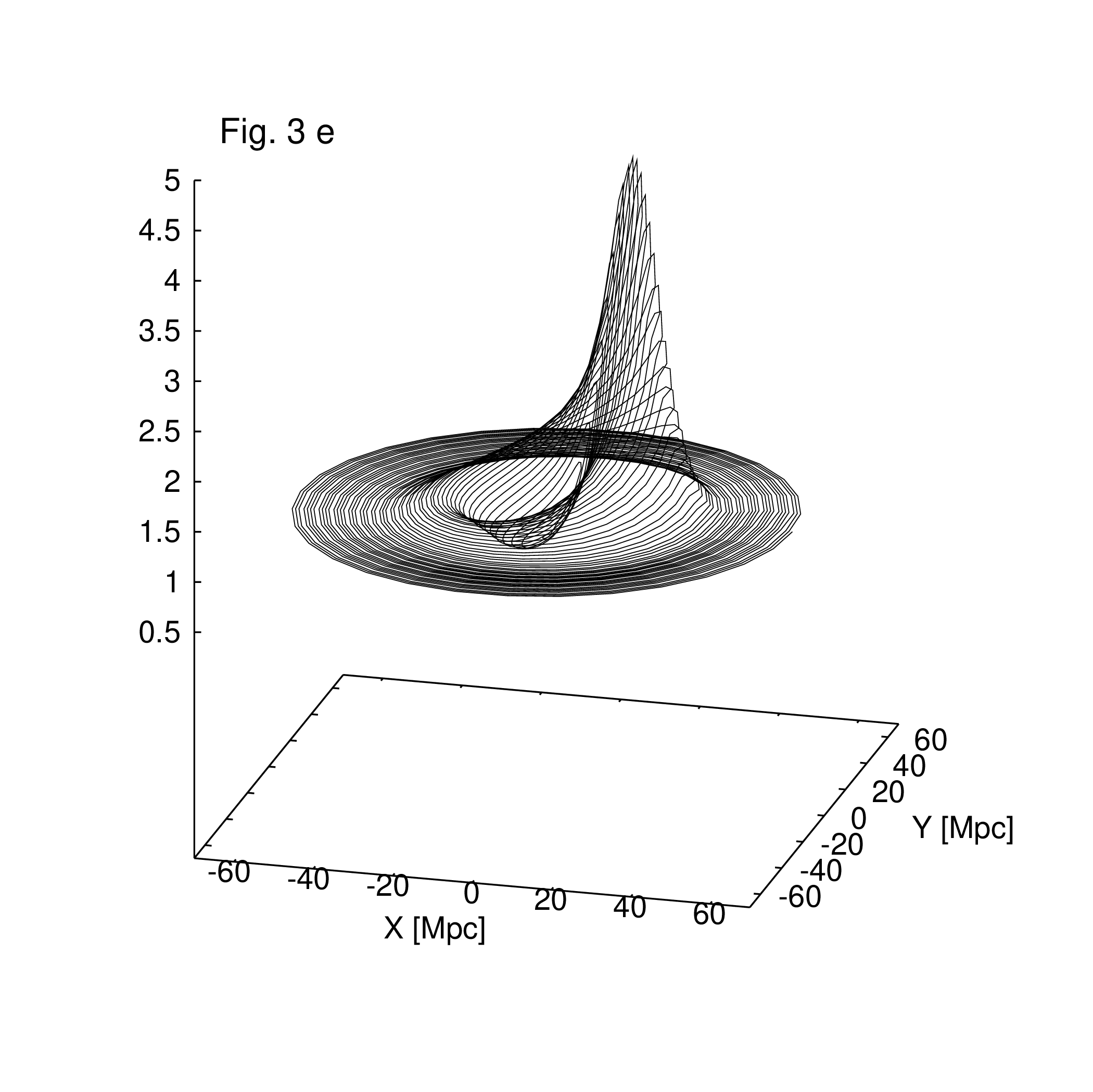}
%%%%%%%%%%%%%%%%%%%%%%%%%%%%%%%%%%%%%%%%%%       
% Colour figs - online only:
%   \includegraphics[scale=0.23]{fig3f_col.eps}
%%%%%%%%%%%%%%%%%%%%%%%%%%%%%%%%%%%%%%%%%%       
% grayscale images - to be published:
   \includegraphics[scale=0.2]{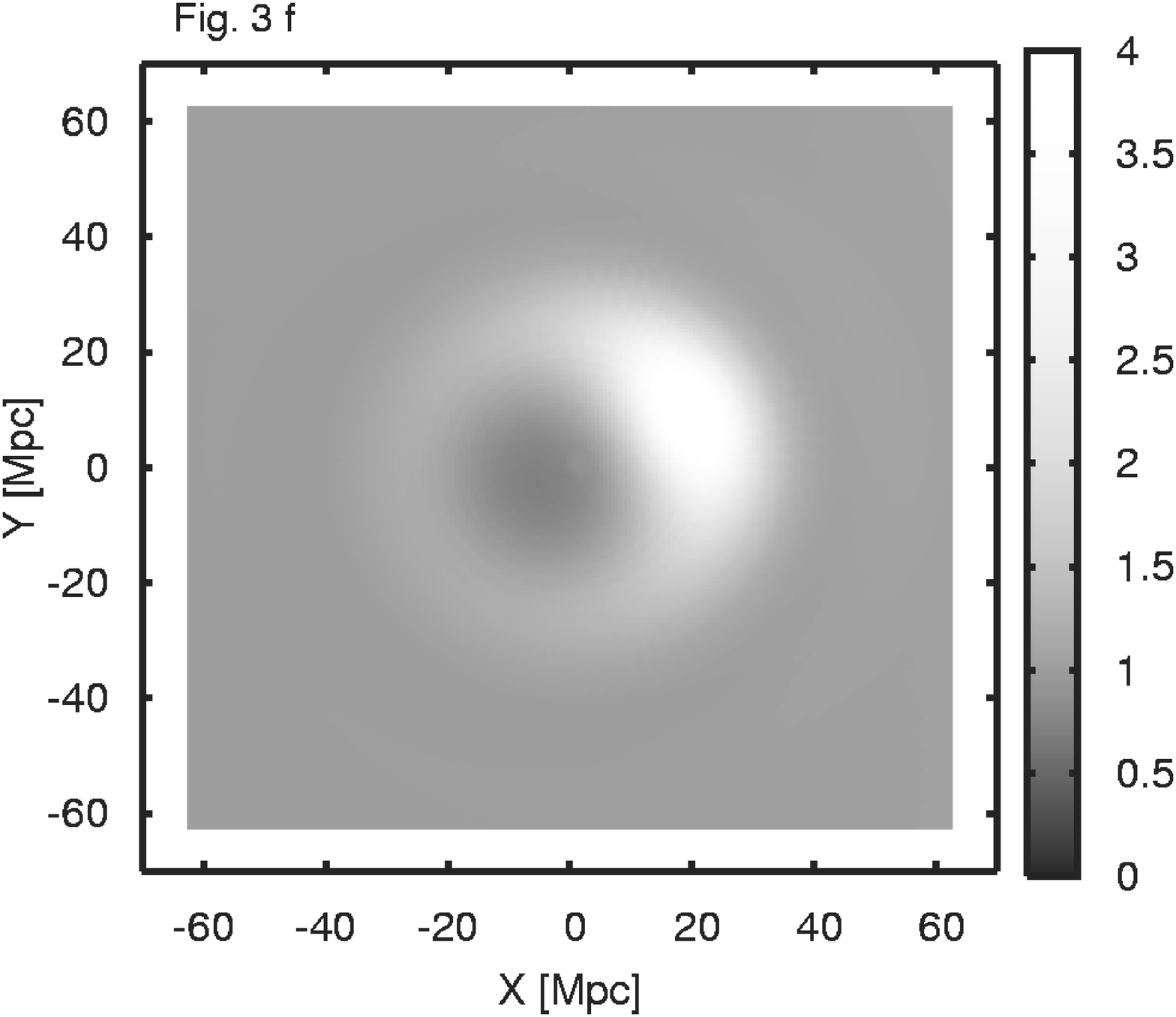}
%%%%%%%%%%%%%%%%%%%%%%%%%%%%%%%%%%%%%%%%%%           
    \caption{
    ("to editor: for online version please use color figures:
    fig3b\_col.eps,fig3d\_col.eps, and fig3f\_col.eps")
     The present day density distribution of model 1 (Sec. \ref{model2}).  
    Fig. 3(a) presents a schematic view. Figs. 3(b) -- 3(f)
     presents the density distribution in background units. 
    Coordinates $X,Y,Z$ are defined as follows:
    $X = \Phi(t_0,r) \sin \theta \cos \phi$,   
    $Y = \Phi(t_0,r) \sin \theta \sin \phi$, $Z = \Phi(t_0,r) \cos \theta$.}
    \label{fig3}
\end{figure*}

The density distribution is presented in Figs.  \ref{fig3}(a) --- \ref{fig3}(f).
Fig. \ref{fig3}(a) presents a schematic view of the stricture. 
Fig \ref{fig3}(b) depicts the vertical cross section, so it goes through the void 
and the cluster [as presented in Fig. \ref{fig3}(a)]. Figs. \ref{fig3}(c)
and \ref{fig3}(d) show the cross section through the equator ($Z = 0$) while
Figs. \ref{fig3}(e) and \ref{fig3}(f) depict the cross section through the 
surface of $Z = - 20$ Mpc [it passes through the cluster presented in 
Fig. \ref{fig3}(b)].

As can be seen, this model does not qualitatively differ from model 1. Both
models present mass distributions similar to dipole structure. It is well known
that the mass distribution in the Szekeres model has the form of a mass-dipole
superposed on a monopole. This was first noticed by Szekeres \cite{Sz2}. The
functions $S$, $P$, and $Q$ simply describe the position of this dipole. As
can be seen from Eq. (\ref{nuz}), the functions $P$, and $Q$ cause that the
density distribution [eq. (\ref{rho})] changes periodically with the period $2
\pi$. Although $\nu,_r$ appears in the denominator as well as in the numerator
of Eq. (\ref{rho}), it is impossible to have the period larger then $2 \pi$
because it would introduce shell crossing singularities (see Hellaby and
Krasi\'nski \cite{HK}; Pleba\'nski and Krasi\'nski \cite{PK} for details on how
to avoid shell crossings in the Szekeres model). The function $S(r)$ on the
other hand, as seen from Eq. (\ref{nuz}) describes the dipole distribution 
along vertical axis. By setting $S$, $P$, and $Q$ constant we drag the dipole 
to the origin and smooth it out to a spherically symmetric mass distribution.

The shell crossing, which was mentioned above,
can also occur during the evolution.
Sometimes it can be avoided by suitable choice of the initial data,
but there are situations when it is impossible
and the Szekeres model breaks down.
This means that pressure cannot be neglected
and a more realistic matter model should be employed.
(It is expected that in those more realistic models, for which no exact
solutions of Einstein's equations are known so far, the shell crossings
would be replaced by regions of high density. These large densities would
become infinite in the limit of zero pressure gradient.)
However, in models presented here, matter density
is not extreme and diameters of considered structures are large,
thus the Szekeres model is appropriate and employing
a more sophisticated model with inhomogeneous pressure distribution is
unnecessary.

\subsection{Evolution}\label{evol}

\begin{figure}
    \begin{center}
    \includegraphics[scale=0.32]{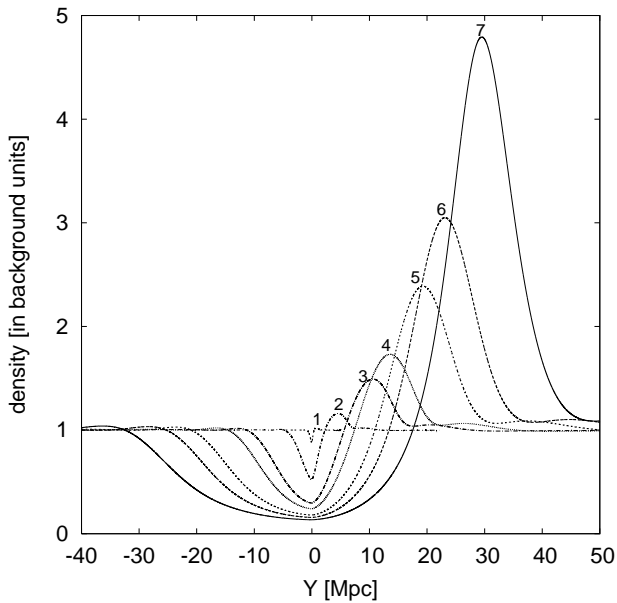}
\caption{The density profile for diffrent time instants: 1 --- 100 My after
the Big Bang, 2 --- 1 Gy,  3 --- 3.5 Gy, 4 --- 5 Gy, 5 --- 8 Gy, 6 --- 10
Gy, 7 --- present instant.}
	\label{fig4}
    \end{center}
\end{figure}

Since model 1 does not differ significantly form the 2, lest us focus only on
the evolution of model 1. The evolution of the model is presented in
Fig. \ref{fig4}. Fig. \ref{fig4} shows the evolution of a density profile which 
goes through the void and the cluster, it is the line $Z = X = 0$ presented
 in Fig. \ref{fig2}(a) --- \ref{fig2}(d). The density distribution is presented 
for different instats, from $100$ million years after the Big Bang 
up to the present.

\subsubsection{Comparison of different approaches}\label{scom}

To estimate how two neighboring structures influence each other's evolution,
let us compare the evolution of the double structure presented above with the
evolution of single structures obtained by other models. The usual way of
calculating the evolution of a density contrast is the the linear approach.
The linear approach is based on the assumption that the density evolves like
in a homogeneous background but with a small correction:

\begin{equation}
\rho(r,t) = \rho_b \left[ 1 + \delta(r,t) \right],
\end{equation}
  Insetring the above formula
into the Einstein equations and after linearising the equations, one obtains:

\begin{equation}
\ddot{\delta} + 2 \frac{\dot{a}_b}{a_b} \dot{\delta} - \frac{1}{2} \kappa c^2
\rho_b \delta = 0
\end{equation}
However, due to the large present density contrast, this approach is in most
cases inadequate. An alternative approach is to use the spherical symmetric
Lema\^itre--Tolman model. Since it is an exact and inhomogeneous solution of
the Einstein equations, one does not have to worry about the smallness of the
present day density contrast. The density contrast is defined similarly as
above:

\begin{equation}
\delta(r,t) = \frac{\rho(r,t) - \rho_b}{\rho_b}.
\end{equation}
The evolution of cosmic structures in the Lema\^itre--Tolman model was studied
in detail by Krasi\'nski and Hellaby \cite{KH1, KH2, KH3}, Bolejko, Krasi\'nski
\& Hellaby \cite{BKH}.

\begin{figure}
    \begin{center}
    \includegraphics[scale=0.24]{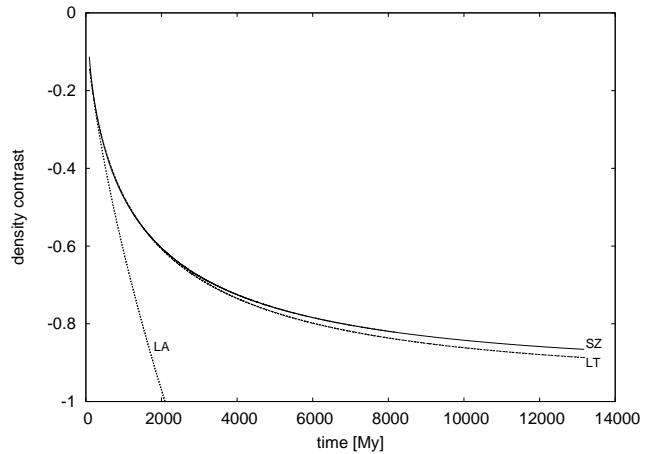}
\caption{The evolution of the density contrast inside the void within the
Szekeres model (SZ), the Lema\^itre--Tolman model (LT) and the linear approach
(LA).}
	\label{fig5}
    \end{center}
\end{figure}

\begin{figure}
    \begin{center}
    \includegraphics[scale=0.24]{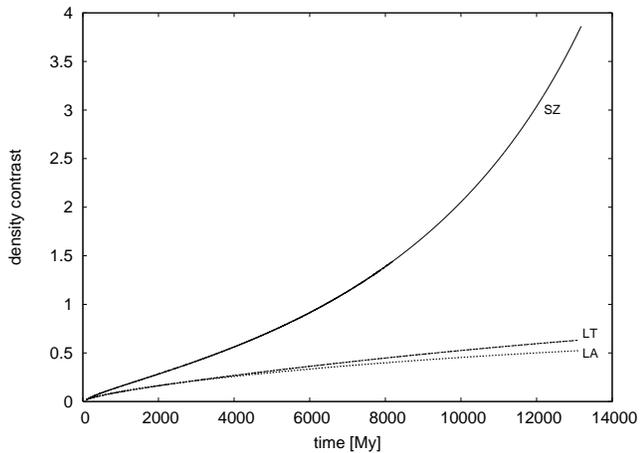}
\caption{The evolution of the density contrast inside the cluster within the
Szekeres model (SZ), the Lema\^itre--Tolman model (LT) and the linear approach
(LA).}
	\label{fig6}
    \end{center}
\end{figure}

The comparison of the evolution of the density contrast in the Szekeres and
Lema\^itre--Tolman models, and in the linear approach is presented in 
Figs. \ref{fig5} and \ref{fig6}. 
These figures present the values of the density contrast at central parts 
of a void (Fig. \ref{fig5}) and a cluster (Fig. \ref{fig6}). 
The initial conditions specifying these models were the same as in the Szekeres 
model. The initial instant was 100 million years after the Big Bang.

Fig. \ref{fig5} presents the evolution of the density contrast inside the void. 
As one can see the linear approach becomes inadequate very soon, and after 
2 Gy it gives unphysical values (the density contrast cannot be smaller 
than $-1$). The evolutions of the density contrast in the Szekeres and 
Lema\^itre--Tolman models are comparable, although the Lemaitre-Tolman 
produces lower values. In the Lema\^itre--Tolman model mass flows from the
central part in all directions with the same rate. In the Szekeres model
the mass-flow depends on the direction, hence this small difference in final 
values of the density contrast. The feature of the mass-flow's direction
 and its significance is more visible in the cluster evolution's case.

Fig. \ref{fig6} presents the evolution of the density contrast inside the
cluster. As one can see, the evolutions of the density contrast in the linear
 approach and Lema\^itre--Tolman model are comparable. The evolution of the 
density contrast in the Szekeres model is significantly faster. This implies 
that the adjourning void plays a significant role in the process of the 
cluster formation. The mass-flow form the void towards the cluster is much 
faster than from other directions. This can be seen as an asymmetry of a void. 
This asymmetry is clearly depicted in Fig \ref{fig4}.

\section{Conclusions}

This paper presents the application of the Szekeres model to the  process of
structure formation. A model of a double structure, i.e. a void with an
adjourning supercluster was constructed. Since this model is based on an exact
solution of Einstein's equations, it presets the evolution of these structures
without such 
 approximations as linearity, hence the interaction 
between described structures can be estimated.

The results show that the mass flow form the void to the cluster enhances the
growth of the density contrast of a galaxy cluster. In the model presented here
the growth of the density contrast was about 5 times faster than in a
spherically symmetric model, and 8 times faster than in the linear approach.
The evolution of the voids is similar to the evolution in the
Lema\^itre--Tolman model but because the spherical models do not distinguish
any direction, the outward mass flow is a little bit faster than in the
Szekeres model. As seen in Figs. \ref{fig5} --- \ref{fig6}, the process of 
the structure formation is a strongly nonlinear process.

The models based on the Szekeres solution have also one more advantage. They
can be used in problems of light propagation, which is impossible in the
N-body simulations. The Szekeres model has still a great, but so far unused,
potential for applications in cosmology, and in the future might be of great
importance in modeling some processes.

\begin{acknowledgments}
I would like to thank Andrzej Krasi\'nski and Charles Hellaby for 
their valuable comments and discusions concerning the Szekeres model. 
Andrzej Krasi\'nski is  gratefully acknowledged for his help with 
preparing the manuscript.
\end{acknowledgments}


\begin{thebibliography}{10}

\bibitem{NB1}
V.~Springel et al. Nature \textbf{435}, 629 (2005)

\bibitem{NB2}
J.~M.~Colberg et al. Mon. Not. R. Astron. Soc., \textbf{319}, 209 (2000)

\bibitem{NB3}
A.~Jenkins et al. Astrophys. J \textbf{499}, 20 (1998)


\bibitem{Sz1}
P.~Szekeres, Commun. Math. Phys. \textbf{41}, 55 (1975a).

\bibitem{BST}
W.~B.~Bonnor, A.~H.~Sulaiman, and N.~Tomimura, Gen. Relativ. Gravit. 
\textbf{8}, 549 (1977).

\bibitem{BT}
W.~B.~Bonnor and N.~Tomimura, Mon. Not. R. Astron. Soc. \textbf{175}, 85 (1976).

\bibitem{Sz2}
P.~Szekeres, Phys. Rev. D \textbf{12}, 2941 (1975b).

\bibitem{HK}
C.~Hellaby and A.~Krasi\'nski, Phys.Rev. D \textbf{66}, 084011 (2002).


\bibitem{Lem}
G.~Lema\^{\i}tre, Ann. Soc. Sci. Bruxelles \textbf{A53}, 51 (1933);
reprinted in Gen. Relativ. Gravit. \textbf{29}, 641 (1997).


\bibitem{Tol}
R.~C.~Tolman, Proc. Nat. Acad. Sci. USA  \textbf{20}, 169 (1934);
reprinted in Gen. Relativ. Gravit. \textbf{29}, 935 (1997).

\bibitem{GW}
S.~W.~Goode and J.~Wainwright, Phys. Rev. D \textbf{26}, 3315 (1982).


\bibitem{NR}
W.~H.~Press, B.~P.~Flannery, S.~A.~Teukolsky, and W.~T.~Vetterling, 
\textit{Numerical Recipes. The art of Scientific Computing}
 (Cambridge Univ. Press, Cambridge, 1986).



\bibitem{Hoy}
F.~Hoyle and M.~S.~Vogeley, Astrophys. J. \textbf{607}, 751 (2004).


\bibitem{Bar}
S.~Bardelli, F.~Zucca, G.~Zamorani, L.~Moscardini, and R.~Scaramella, 
Mon. Not. R. Astron. Soc.
\textbf{312}, 540 (2000).

\bibitem{Kol}
T.~Kolatt, A.~Dekel, and O.~Lahav, Mon. Not. R. Astron. Soc. 
\textbf{275}, 797 (1995).


\bibitem{Hud}
M.~J.~Hudson, Mon. Not. R. Astron. Soc. \textbf{265}, 43 (1993).


\bibitem{PK}
J.~Pleba\'nski and A.~Krasi\'nski,  \textit{Introduction to general relativity 
and cosmology} (Cambridge Univ. Press, Cambridge, in press).


\bibitem{KH1}
A.~Krasi\'nski and C.~Hellaby, Phys. Rev. D \textbf{65}, 023501 (2002).

\bibitem{KH2}
A.~Krasi\'nski and C.~Hellaby, Phys. Rev. D  \textbf{69}, 023502 (2004).

\bibitem{KH3}
A.~Krasi\'nski and C.~Hellaby, Phys. Rev. D  \textbf{69}, 043502 (2004).


\bibitem{BKH}
K.~Bolejko, A.~Krasi\'nski and C.~Hellaby, Mon. Not. R. Astron. Soc. 
\textbf{362}, 213 (2005).


\end{thebibliography}
\end{document}